\documentclass[graybox,natbib,nosecnum]{svmult}

\usepackage{mathptmx}       % selects Times Roman as basic font
\usepackage{helvet}         % selects Helvetica as sans-serif font
\usepackage{courier}        % selects Courier as typewriter font
\usepackage{type1cm}        % activate if the above 3 fonts are

\usepackage{makeidx}         % allows index generation
\usepackage{graphicx}        % standard LaTeX graphics tool
\usepackage{multicol}        % used for the two-column index
\usepackage[bottom]{footmisc}% places footnotes at page bottom
\usepackage[normalem]{ulem}  % for strike-through of text with \sout{}  
\usepackage{hyperref}        %for hyperlinks

\makeatletter
\providecommand*{\toclevel@titlech}{0}
\providecommand*{\toclevel@authorch}{0}
\makeatother

\newcommand*\aap{A\&A}

\newcommand*\aapr{A\&A Rev}

\newcommand*\aj{AJ}

\newcommand*\apj{ApJ}
\newcommand*\apjl{ApJ}

\newcommand*\apss{Ap\&SS}
\newcommand*\araa{ARA\&A}

\newcommand*\baas{BAAS}

\newcommand*\grl{Geophys Res Lett}

\newcommand*\icarus{Icarus}

\newcommand*\jgr{J Geophys Res}

\newcommand*\mnras{MNRAS}

\newcommand*\nat{Nature}

\newcommand*\planss{Planet Space Sci}

\newcommand*\psj{Planet Sci J}

\newcommand*\ssr{Space Sci Rev}

\newcommand{\hbindex}[1]{{#1}\index{#1}}

\newcommand\lesssim{\mathrel{\hbox{\rlap{\hbox{\lower4pt\hbox{$\sim$}}}\hbox{$<$}}}}
\newcommand\gtrsim{\mathrel{\hbox{\rlap{\hbox{\lower4pt\hbox{$\sim$}}}\hbox{$>$}}}}

\makeindex             % used for the subject index
\begin{document}
\title{Radio Observations as an Extrasolar Planet Discovery and Characterization: Interior Structure and Habitability}

\titlerunning{Exoplanets at Radio Wavelengths}
\author{T.~Joseph~W.~Lazio}

\institute{Jet Propulsion Laboratory, California Institute of
  Technology, 4800 Oak Grove Dr., M/S~67-201, Pasadena, CA  91109,
  \hbox{USA}, \email{Joseph.Lazio@jpl.nasa.gov};
  \copyright 2023.  Government sponsorship acknowledged.
}

\maketitle

\abstract{%
  Detection of radio emission from Jupiter was identified quickly as
  being due to its planetary-scale magnetic field.  Subsequent
  spacecraft investigations have revealed that many of the planets,
  and even some moons, either have or have had large-scale magnetic
  fields.  In the case of the Earth, Jupiter, Saturn, Uranus, and
  Neptune, their magnetic fields are generated by dynamo processes
  within these planets, and an interaction between the solar wind and
  their magnetic fields generates intense radio emission via the
  electron cyclotron maser instability.  In the case of Jupiter, its
  magnetic field interacts with the moon Io to result in radio
  emission as well.
  Extrasolar planets reasonably may be expected to generate
  large-scale magnetic fields and to sustain an electron cyclotron
  maser instability.  Not only may these radio emissions be a means
  for discovering extrasolar planets, because magnetic fields are tied
  to the properties of planetary interiors, radio emissions may be a
  remote sensing means of constraining extrasolar planetary properties
  that will be otherwise difficult to access.  In the case of
  terrestrial planets, the presence or absence of a magnetic field may
  be an indicator for habitability.  Since the first edition of the
  Handbook, there have been a number of advances, albeit there remain
  no unambiguous detection of radio emission from  extrasolar planets.  
  New ground-based telescopes and new possibilities for space-based
  telescopes provide promise for the future.}

\clearpage

\section{Introduction}\label{sec:intro}

While the planet Jupiter has been known since antiquity, the
serendipitous discovery of its decametric wavelength ($\lambda \sim
10\,\mathrm{m}$, $\nu \sim 30\,\mathrm{MHz}$) radio emission
\citep{bf55,fb56} illustrates how planetary radio emissions could be
used to detect (extrasolar) planets.  Indeed, one of the developments
since the first edition of the Handbook was published has been
publications reporting possible extrasolar planetary analogs.
(See ``Observational Constraints for Extrasolar Planets'' below.)

Jupiter's decametric radio emission is linked to its magnetic field
\citep[and references within]{cg69}, and, in the Solar System, the
combination of decametric radio emissions, other remote sensing
measurements, and \textit{in situ} spacecraft measurements have
established that the Earth, all of the giant planets, Mercury, and
Jupiter's moon Ganymede contain internal \hbindex{dynamo} currents
that generate planetary-scale magnetic fields.

Since the detection of Jupiter's radio emission, Earth and
the other three giant planets in the Solar System, Saturn, Uranus, and
Neptune, all have been confirmed to emit at radio wavelengths via a similar process
\citep{z92}:
Immersed in the stellar wind of a host star, a planetary-scale
magnetic field forms a \hbindex{magnetosphere} that can extract energy
from the stellar wind.  Some of this energy then can be radiated via
the \hbindex{electron cyclotron maser instability}, likely at
decametric wavelengths or longer.
Notably, in the case of Uranus and Neptune, their luminosities were
\emph{predicted} before the Voyager~2 encounters \citep{dk84,d88,mg88}.

\begin{figure}[tbh]
  \centering
  \includegraphics[width=0.47\textwidth]{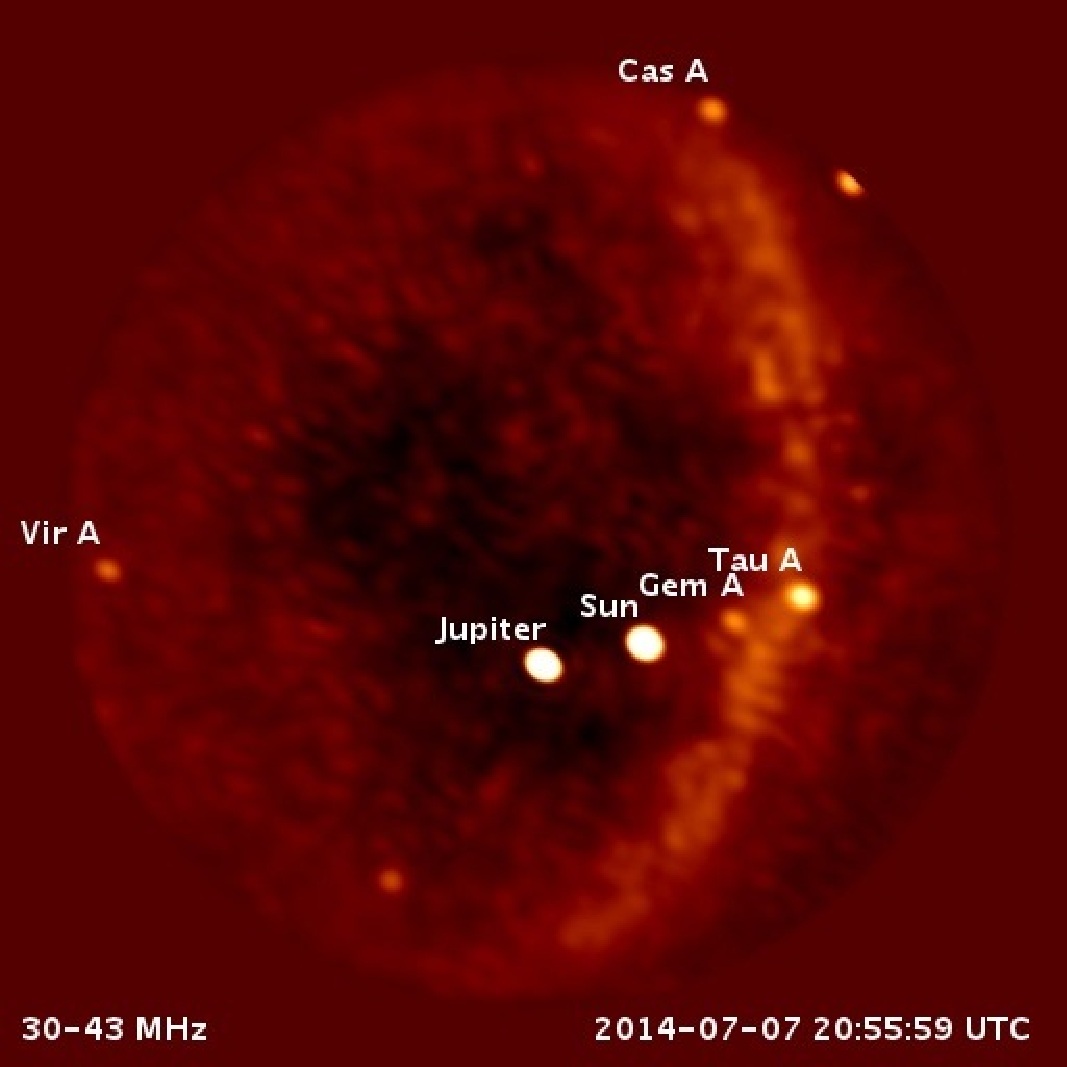}\,\hfil%
  \includegraphics[width=0.47\textwidth]{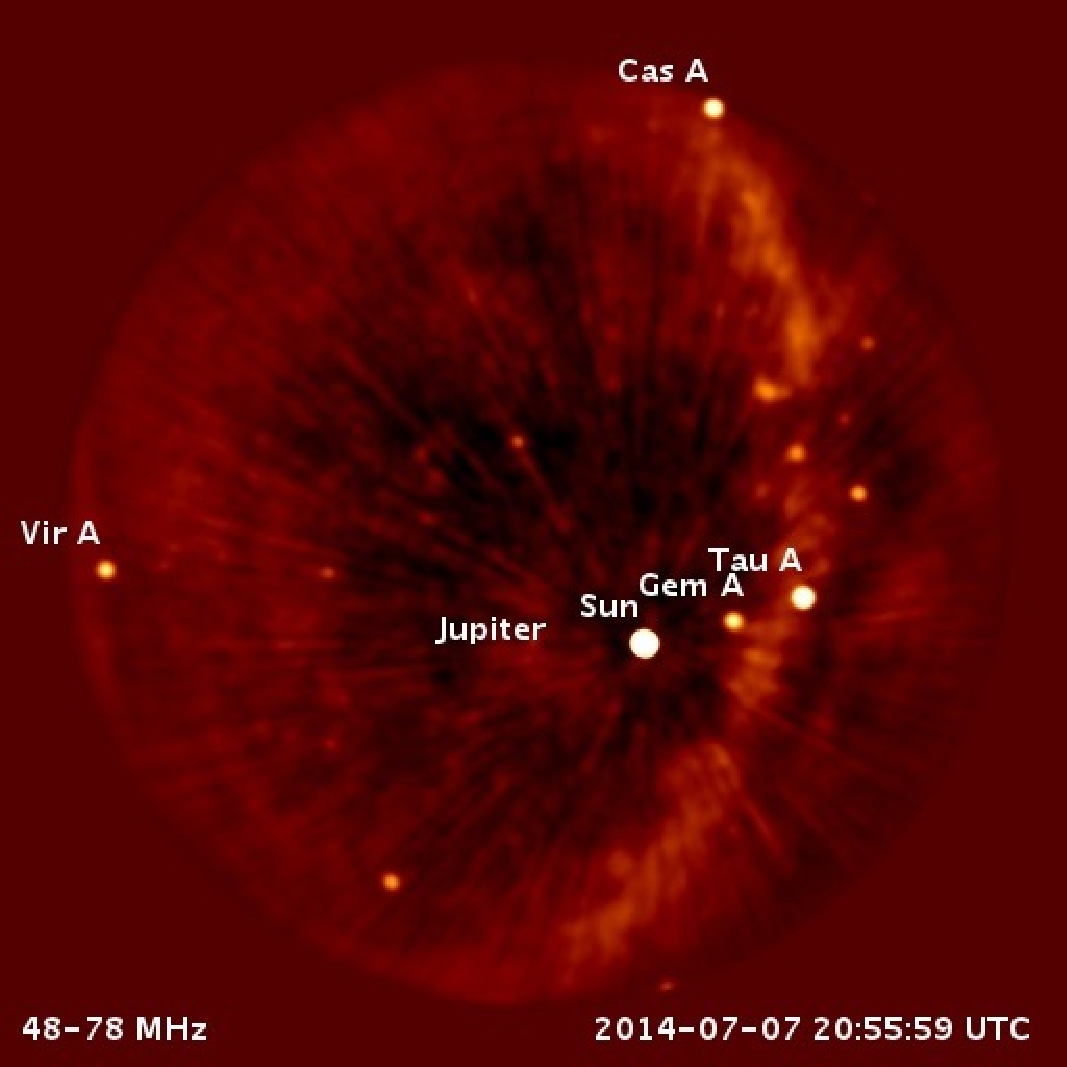}
  \vspace*{-1ex}
  \caption[Jupiter and the Frequency Truncation of the Electron
    Cyclotron Maser Instability]%
          {Radio sky in the 30~MHz--43~MHz (\textit{left}) and
            47~MHz--78~MHz bands (\textit{right}); zenith is at the
            center of the images.  Strong sources are labeled, notably
            including Jupiter and the Sun.  In the lower frequency
            image, Jupiter is of comparable brightness to the Sun,
            illustrating that the star-planet intensity ratio can be
            of order unity at radio wavelengths.  In the higher
            frequency image, the absence of Jupiter is consistent with the
            exceptionally strong cutoff of \protect\hbindex{cyclotron
              maser emission} where the local plasma frequency exceeds
            the local cyclotron frequency within the planet's
            \protect\hbindex{magnetosphere}.  (Images courtesy of
            M.~Anderson)}
          \label{fig:jupiter}
\end{figure}

Even before the confirmation of extrasolar planets, extrapolation from
the planetary radio emissions in the Solar System led multiple groups
to attempt to detect analogous emissions, typically by targeting
nearby stars in hopes that they would be orbited by one or more giant
planets \citep{yse77,wdb86}.
Indeed, in vivid contrast to the case at other wavelengths, the
star-planet ratio for radio emissions can be of order unity, and
during its most intense radio bursts, Jupiter even can be
\emph{brighter} than the (quiet) Sun (viz.\ Figure~\ref{fig:jupiter}).

Extrapolated to extrasolar planets, radio emissions provide not only a
potential means for discovering or detecting planets, but they would
provide a direct measure of an extrasolar planet's magnetic field, in
turn placing constraints on the thermal state, composition, and
dynamics of its interior---all of which will be difficult to determine
by other means.
In the case of a terrestrial planet, the detection of a magnetic field
also may provide crucial information about the extent to which its
surface is shielded from energetic particles and potentially
\hbindex{habitable}.

This chapter reviews how radio emissions could be used not only to
discover extrasolar planets, but also how they might be used to study
extrasolar planets, regardless of the means of discovery.
Since the first version of the Handbook, there have been a number of
developments in searches for magnetically-generated radio emission,
potential magnetic field-moderated star-planet interactions, advances
in theoretical understandings about planetary magnetic fields, and
constraints on planetary magnetic fields from other techniques.
\cite{lorentz} provide a recent, complementary review of some of these
topics, extending to cover radio emission from brown dwarfs and stars
as well.

The structure of this chapter is the following.
The next section, ``Observational Constraints for Extrasolar
Planets,'' presents a summary of efforts to date, focussing on recent
results.
The following four sections---``Planetary Magnetic Fields,'' ``The
Electron Cyclotron Maser Instability and (Extrasolar) Planetary Radio
Emission,'' ``Observational Considerations,'' and ``Extensions to and
Predictions for Extrasolar Planets''---are intended to be high-level
summaries of fundamental aspects motivating observations at radio
wavelengths.  Some of material in these initial sections also is
relevant for describing the radio emission from brown dwarfs, a topic
discussed in brief in the final section.
The following sections---``Planetary Magnetic Fields and Interiors,''
and ``Planetary Magnetic Fields and Habitability''---present the
implications of detection of magnetically-generated radio emission
from planets.  These sections have been revised substantially
since the first version of this chapter to take into account the
latest developments.
A concluding section, ``Future Steps,'' envisions likely progress in
the next decade and beyond.

The focus of this chapter is on \emph{\hbindex{electron cyclotron maser
  instability}} (ECMI) emissions resulting from a stellar wind-planetary
\hbindex{magnetosphere} interaction.
While planets, or at least Jupiter, can
generate synchrotron emission that appears at decimeter and
centimeter wavelengths, it is likely to remain well beyond detection
capabilities in the near future.
As an example, the centimeter-wavelength flux density of Jupiter is of
order a few Janskys \citep[e.g.,][]{ktb89,dpbg+03}.  Scaled to a
distance of even a few parsecs, the resulting flux density would be
\emph{sub-nanoJansky}, which is well beyond even the most optimistic
expectations for next-generation centimeter-wavelength telescopes,
such as the Phase~1 of the Square Kilometre Array 
\citep[SKA1, cf.][]{cwb22}.
Planets, or more likely proto-planets, likely also are to be detected
at centimeter to millimeter wavelengths \citep[e.g.,][]{w08,pca+16},
with the the Very Large Array (VLA) or the Atacama Large
Millimeter/submillimeter Array (ALMA), and there is increased
potential of such studies with the next-generation Very Large Array
(ngVLA).  Clearly promising, but also poised for significant
discoveries over the next few years, this chapter defers any
discussion of radio emission at these wavelengths.  Finally, this
chapter does not consider the possibility of detecting artificial
transmissions, though this approach may represent the ultimate in
extrasolar planet discovery \citep{t01}.

\section{Observational Constraints for Extrasolar Planets}\label{sec:exoobserve}

Since the first edition of the Handbook, there have been a number of
notable developments, including multiple claims of detections of
extrasolar planets at radio wavelengths.

\begin{figure}[thb]
\centering
\includegraphics[angle=-90,width=0.98\textwidth]{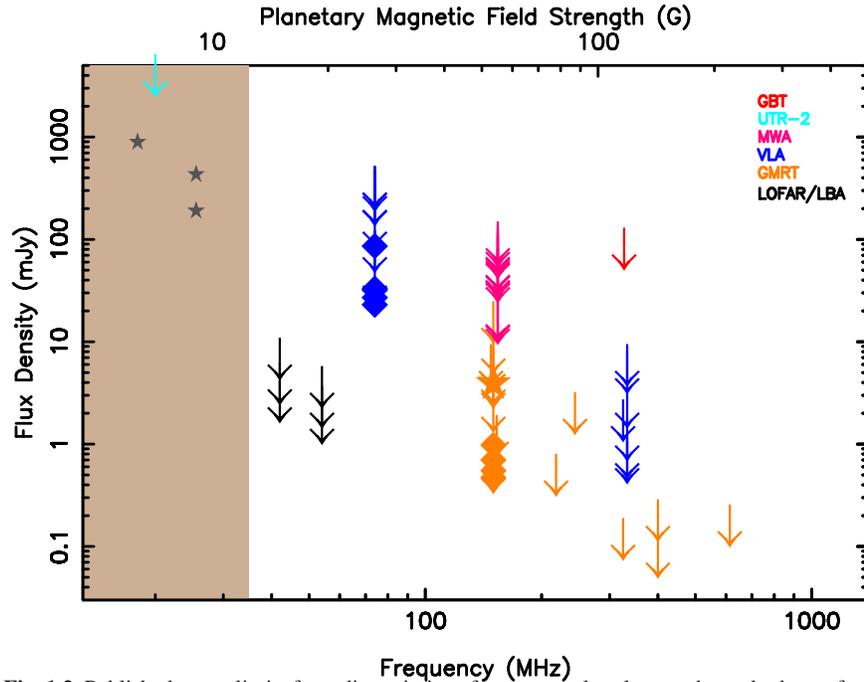}
\vspace*{-3ex}
\caption[Radio Emission Limits for Extrasolar Planets]%
{Published upper limits for radio emissions from extrasolar planets,
  shown both as a function of radio frequency (bottom axis) and
  implied planetary magnetic field strength (top axis).  Observations
  have been obtained by, and are color-coded for, the Ukranian
  T-shaped Radio Telescope (UTR-2), Green Bank Telescope (GBT),
  Murchison Widefield Array (MWA), Very Large Array (VLA), Giant
  Metrewave Radio Telescope (GMRT), and the Low Frequency
  Array/Low-Band Antennas (LOFAR/LBA).  For clarity, the upper limits
  at~150~MHz on all extrasolar planets known at the time and in the
  footprint of the Tata Institute of Fundamental Research (TIFR)-GMRT
  Sky Survey (TGSS) are not shown \citep{sldeg-kki-c14}, and only the
  ensemble analysis results of \cite{2022ApJ...926..228L} are shown.
  Solid
  diamonds show  statistical limits on the average
  planetary radio emission from various samples of planets
  \citep{l+10,2022ApJ...926..228L}.  The solid stars represent the tentative
  detection of HAT-P-11b at~150~MHz \citep{ldesg-kz13} and the
  unconfirmed detections of $\tau$~Boo~b near~20~MHz
  \citep{2021A&A...645A..59T,tzg+23}.  The shaded region shows the
  approximate frequency range for Jovian decametric emissions
  (cf.\ Figure~\ref{fig:radiosolarsystem}).  All of these observations
  have been from ground-based telescopes, emphasizing that future
  space-based observations likely will be needed to study lower-mass
  planets.}
\label{fig:radio_limits}
\end{figure}

\begin{itemize}
\item%
  \cite{2021A&A...645A..59T} reported a detection of radio emission
  from~$\tau$~Boo~b at a frequency near~25~MHz.
  From the earliest quantitative estimates for extrasolar planetary
  radio emission \citep{fdz99,lf+04,gzs07}, this planet has
  been identified consistently as likely to have emissions both strong
  enough and at a high enough frequency to be detectable from the
  ground.  However, a notable aspect of the reported radio emission was
  that it was relatively narrow band ($\Delta\nu/\nu \sim 0.3$), in
  contrast to the emission from Jupiter for which $\Delta\nu/\nu \approx
  1$.  Unfortunately, subsequent efforts to detect this emission
  from~$\tau$~Boo~b have not been successful \citep{tzg+23}.

\item%
  Substantial progress has been achieved in calibration of Low
  Frequency Array (\hbindex{LOFAR}) observations at~50~MHz, with the
  result that the upper limits on the radio emissions from HD~80606b
  are at the level of a few millijanskys \citep{2020A&A...644A.157D}.
  As Jupiter emits up to approximately 35~MHz, this improved
  capability means that searches can be conducted at frequencies
  comparable to those at which Jupiter emits and with sensitivities
  consistent with what might be expected for nearby extrasolar
  planets.  Unfortunately, HD~80606b itself is approximately 60~pc
  distant.  Even the most optimistic projections for its radio
  emission suggest that it is likely an order of magnitude fainter
  than the current limits \citep{2020A&A...644A.157D}.

\item%
  There have been multiple reports of low-mass stars producing radio
  emission, including for the nearby M~dwarf GJ~1151 \citep{vcs+20},
  the flare star AD~Leo \citep{2023arXiv230600895Z}, and YZ~Ceti
  \citep{pv23,tbl+23}.  For at least some of the stars, the radio
  emission is the result of the \hbox{ECMI}, notably AD~Leo, and there
  have been multiple claims of magnetic star-planet interactions,
  including estimates of planetary magnetic field strengths (0.4~G in
  the case of YZ~Ceti).  However, in the case of GJ~1151, while
  \cite{2023arXiv230104442B} detected a planet orbiting GJ~1151, its
  mass is larger than that argued initially to be responsible for the
  radio emission from that system, and, in all cases, ambiguity
  remains about whether the star is emitting or the planet is
  emitting.  Finally, as noted below, a unipolar induction mechanism
  can generate radio emissions, as in the Jupiter-Io system.  As such,
  while the detection of radio emissions could provide evidence for
  the existence of a planet, in general, little information about its properties
  may result.

\item%
  There have been searches focussed on detecting the extrasolar
  planetary equivalent of Jupiter's emission that is driven by the
  interaction between its \hbindex{magnetosphere} and its moon Io
  \citep{2023AJ....165....1N,2023MNRAS.522.1662N}.

\item%
  Multiple observations have been conducted of M~dwarfs and brown
  dwarfs, including of TRAPPIST-1, which have shown that they can produce
  coherent radio emission in the absence of a planet or that the
  presence of planets does not lead necessarily to magnetic
  star-planet interactions generating radio emissions
  \citep{2017ApJ...846...75P,ljwhm18,ph18,vh19,cpf+21,cvs+21}.
\end{itemize}

Figure~\ref{fig:radio_limits} presents a graphical summary of most
published limits on the radio emission from extrasolar planets
\citep{zqr+97,bdl00,lf+04,gs07,lf07,ldesg-kz09,sccgjlb09,l+10,ldesg-kz11,ldesg-kz13,hsa+13,sldeg-kki-c14,mbk+14,2020A&A...644A.157D,gm21,2022ApJ...926..228L,2023AJ....165....1N,2023MNRAS.522.1662N,2023MNRAS.tmp.3811S}.
Not shown are a few observations at frequencies above~1000~MHz and a
few observations at frequencies around~20~MHz---limits above~1000~MHz
are not shown as these are likely to be at too high of a frequency,
and would require planetary magnetic field strengths larger than
500~\hbox{G}, while the published limits around~20~MHz are typically
above the range of flux densities shown or do not have adequate
information to assess or both.  Based on a number of predictions that
it is a promising target for detection, the most intensively studied
planet to date is $\tau$~Boo~b.

Several items deserve mention.  First, it is apparent that most
searches have been conducted at frequencies sufficiently high that
Jupiter would not have been detected.  Of all planets in the solar
neighborhood, it is unlikely that Jupiter has the strongest magnetic
field (cf.\ eqn.~[\ref{eqn:fmax}]).  Nonetheless, at least some of the
non-detections plausibly can be attributed to searching at frequencies
that are too high.

Second, the trend of upper limits becoming less constraining at lower
frequencies is real and represents limits on radio telescope
sensitivities at these frequencies.  A primary factor determining the
telescope sensitivity is $A_{\mathrm{eff}}/T_{\mathrm{sys}}$, the
ratio between the effective area of the telescope and the system
temperature.  For a given telescope (e.g., VLA or GMRT), the effective
area~$A_{\mathrm{eff}}$ is essentially fixed (by the number of
antennas and the diameter of each antenna).  At these frequencies, the
dominant contribution to the system temperature~$T_{\mathrm{sys}}$ is
the sky temperature or the power contributed by the Milky Way Galaxy's
synchrotron radiation.  This temperature increases dramatically to
lower frequencies, scaling approximately with frequency as
$\nu^{-2.6}$ \citep{c79}
Consequently, the limits become less constraining at lower
frequencies.  For a dipole-based array (see below), 
the effective area of the individual dipoles scales with frequency
as approximately $\nu^{-2}$, so that any limits that they place should
be much more constant with frequency.

Third, the solid diamond at~74~MHz is the upper limit on the average
planetary radio emission from planets orbiting nearby solar-type stars
\citep{l+10}.  It was constructed from a stacking analysis of the
radio emission in the direction of stars within~40~pc.  As such, it
represents a limit on the combination of the average planetary radio
luminosity and the fraction of solar-type stars hosting planets that
radiate at~74~MHz.

Fourth, the solid star at~150~MHz is the tentative detection, on a
single day, of radio emission from HAT-P-11b \citep{ldesg-kz13}.  If
this measurement represents an actual detection, it implies that the
magnetic field strength of HAT-P-11b is 50~\hbox{G}.  However, equally
sensitive observations on another day did not detect any radio
emission.  Thus, as even the authors acknowledge, some caution is
warranted in concluding that this measurement represents the first
discovery of extrasolar planetary radio emission.

Fifth, as noted above, any claim of detection of the
radio emission from an extrasolar planet must address whether it the
extrasolar planet or host star has been detected.  In general,
extrasolar planetary radio emission is expected to exceed the emission
of the planetary host star \citep{zqr+97,gmmr05}.  Further approaches
suggested to distinguish planetary from stellar radio emission have
included whether the radio emission is modulated with the planet's
orbital period (if known) or with a time scale characteristic of Solar
System planetary rotational periods ($\approx 10$~hr).  However,
\cite{2010MNRAS.406..409F} note that even a purely planetary signal
may be partially modulated by the stellar rotation period, which could
complicate the discrimination between a stellar and a planetary	radio
signal.

Finally, nearly all, if not all, of these searches have targeted known
extrasolar planets.  Only recently, with blind surveys being conducted at
\hbindex{LOFAR} \citep[e.g.,][]{vcs+20,v-lotss}
would the discovery of new extrasolar planets be possible.

\section{Planetary Magnetic Fields}\label{sec:bfields}

There is a rich literature on the topic of the generation and
sustainment of planetary magnetic fields, and a full review is beyond
the scope of this chapter.  Interested readers are encouraged to
consult \cite{Christensen10}, \cite{Stevenson2010SSR}, and
\cite{Schubert&Soderlund2011PEPI} for extensive reviews and
\cite{chr09} for extension to extrasolar planets and dwarf stars; a
W.~M.~Keck Institute for Space Studies Program on Planetary Magnetic
Fields also considered both the state of knowledge of the field and
observational manifestations of planetary magnetic fields, beyond
likely radio detections \citep{kiss}.  Nonetheless, a brief
consideration of planetary magnetic fields is essential for this
chapter, both because they are a critical element for the generation
of radio emission and because their presence may provide information
on extrasolar planets (and Solar System planets?) that will be
difficult to obtain through any other means.

Within the Solar System, a diversity of magnetic field configurations
and amplitudes are observed.  They are observed on planets, some small
bodies, and some moons.  Configurations range from predominantly
dipolar to asymmetric but large-scale to probably residual crustal
magnetism.  Amplitudes range from more than 10~G in strength (Jupiter)
to much weaker and even induced fields (e.g., Io, Europa, Callisto).

\begin{figure}[tbh]
\centering
\includegraphics[width=0.9\textwidth]{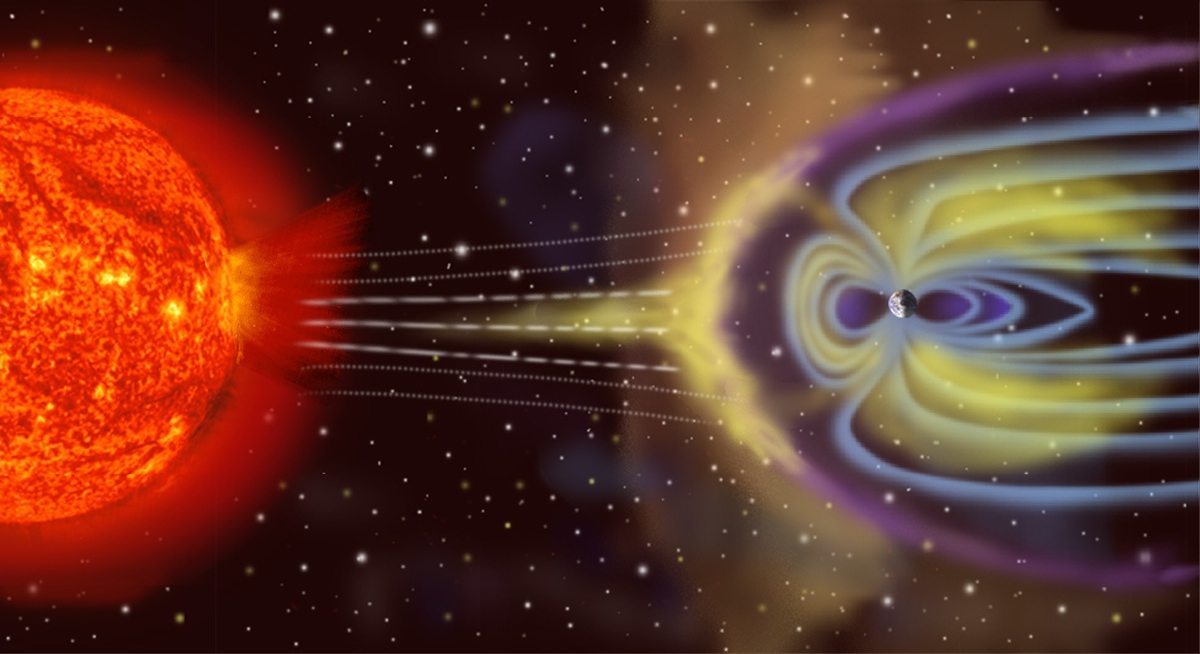}
\vspace*{-1ex}
\caption[Planetary Magnetosphere]%
{Artist's impression of Earth's \protect\hbindex{magnetosphere}, which is produced by
  the interaction between the solar wind and Earth's magnetic field.
  The Earth's magnetic field is generated by internal \protect\hbindex{dynamo} currents,
  and the solar wind-\protect\hbindex{magnetosphere} interaction both shields the
  Earth's atmosphere and surface and produces intense radio emission
  from its polar region via the electron cyclotron maser instability.
  Similar processes occur at Jupiter, Saturn, Uranus, and Neptune,
  though the natures of their dynamos are different than that of
  Earth, and the resulting radio emissions should be detectable over
  interstellar distances.  (Image credit: NASA)}
\label{fig:magnetosphere}
\end{figure}

Of relevance to detection over interstellar distances, it is only the
planets having magnetic fields with strengths of order 1~G and
larger---Earth, Jupiter, Saturn, Uranus, and Neptune---that are
capable of generating a planetary-scale \hbindex{magnetosphere}
(Figure~\ref{fig:magnetosphere}) and therefore sufficiently strong
radio emissions.

It is nearly certain that the generation of these planetary-scale
magnetic fields requires a \hbindex{dynamo} process, in which kinetic
energy in a conducting medium is converted into a magnetic field.  The
\hbindex{dynamo} currents interior to a planet may arise from
differential rotation, convection, compositional dynamics, or a
combination of these processes.  At the very least, it is clear that
the conducting fluids within the different ``magnetic planets'' are
different, with Earth containing liquid iron in
its (outer) core, Jupiter and Saturn likely having metallic hydrogen,
and Uranus and Neptune likely due to salty water under high pressure.
The topic of planetary dynamos is revisited in the section ``Planetary
Magnetic Fields and Interiors'' to discuss both how radio wavelength
emissions might provide constraints on extrasolar planetary interiors
and the potential diversity of planetary interiors.

\section{The Electron Cyclotron Maser Instability and (Extrasolar)
	Planetary Radio Emission}\label{sec:radio}

All of the ``radio active'' planets in the Solar System (Earth,
Jupiter, Saturn, Uranus, and Neptune) produce radio emission via the
\emph{\hbindex{electron cyclotron maser instability}} (ECMI).  This section
provides an introduction to the conditions necessary to create the
\hbox{ECMI};  \cite{t06}, \cite{vkc+11}, and \cite{bt22} provide extensive
discussions of the \hbox{ECMI}.

A heuristic summary of the ECMI as applied to radio emission from
planetary \hbindex{magnetospheres} is to consider a (monoenergetic)
beam of electrons traveling along magnetic field lines.  The electrons
will gyrate around the field lines, producing cyclotron radiation at
the electron \emph{cyclotron frequency}~$\Omega_{\mathrm{ce}}$.  (Ions
also gyrate around magnetic field lines at a
frequency~$\Omega_{\mathrm{ci}}$, but the (much) larger mass of ions
means that this frequency is sufficiently low that it is not relevant
for this discussion.)  If the local environment were a vacuum, this
cyclotron radiation would escape to infinity, but any realistic
\hbindex{magnetosphere} has an ambient plasma.  A plasma has a
characteristic \emph{plasma frequency}~$\omega_{\mathrm{pe}}$,
determined by the (local) plasma density, $\omega_{\mathrm{pe}}
\propto \sqrt{n_e}$.  In general, radiation with a frequency~$\nu$
will neither propagate through nor escape from a region for which $\nu
< \omega_{\mathrm{pe}}/2\pi$.  The fact that the Solar System's
``radio active'' planets produce detectable ECMI emissions indicates
that planetary \hbindex{magnetospheres} reasonably can be expected to
sustain conditions such that $\nu \sim \Omega_{ce} >
\omega_{pe}/2\pi$.

Two effects can combine to enable the escape of cyclotron radiation
from a planetary \hbindex{magnetosphere}.  First, in a low density environment,
the plasma frequency can approach or even be less than that of the
cyclotron frequency, $\omega_{\mathrm{pe}} \lesssim
\Omega_{\mathrm{ce}}$.  Second, the velocity of the electrons
introduces a Doppler shift.  If the electron velocity is high enough,
the electron cyclotron radiation is shifted into resonance with
radiation modes that can escape the planetary \hbindex{magnetosphere} and the
planet radiates in the radio.

In a realistic \hbindex{magnetosphere}, the electrons will not be
monoenergetic but will have some distributions of energies.  If the
electron energy distribution has more electrons with energies above
the escaping cyclotron modes, the electrons will feed energy into the
escaping cyclotron radiation modes, making the emission even more
intense.  By analogy to the inverted population states that can give
rise to laser and maser emission, having an electron energy
distribution that ``stimulates'' more intense radiation is termed a
``maser instability,'' leading to the \hbox{ECMI}.  In general, the
ECMI requires an electron energy distribution with a supra-thermal, 
non-Maxwellian component---for a Maxwellian energy distribution always
has more electrons at lower energies.

A more exact treatment would consider the dispersion relation of waves
in a magnetized plasma, but essentially the same conclusions would be
reached.  A population of energized electrons traveling along magnetic
field lines through a low plasma density environment can generate
intense radio emissions via a resonance with the (Doppler shifted)
electron cyclotron frequency.

\section{Observational Considerations}\label{sec:observe}

Several observationally-relevant conclusions can be drawn, either from
the heuristic approach described in the previous section or from a
more exact treatment.  First, ECMI planetary radio emission should be
highly circularly polarized.  This conclusion follows simply from the
nature of cyclotron emission itself (produced by gyrations around
magnetic field lines), and circularly polarized emission is the
standard for planetary radio emissions in the Solar System.  As 
few radio astronomical sources are circularly polarized, searching for
circularly-polarized radio sources is a common technique for
identifying candidate extrasolar planets (or stars) emitting via the
ECMI \citep{v-lotss}.

Second, a planet will radiate up to a maximum (radio) frequency
determined by the largest magnetic field strength within the region
where the conditions for the ECMI can be sustained \citep{fdz99}.  In
practice, this region is typically near the magnetic polar regions,
for which the maximum radiated frequency is then
\begin{eqnarray}
\nu_{\mathrm{max}}
 & =     & \frac{e{\cal{M}}R_p^3}{2\pi m_e} = \frac{eB_{\mathrm{pole}}}{2\pi m_e},\nonumber\\
 &\approx& 2.8\,\mathrm{MHz}\left(\frac{B_{\mathrm{pole}}}{1\,\mathrm{G}}\right),
\label{eqn:fmax}
\end{eqnarray}
where $e$ is the charge on the electron, $m_e$ is the mass of the
electron, $R_p$ is the radius of the planet, $\cal{M}$ is the magnetic
moment of the planet at the surface or cloud tops (as distinct from
the magnetic moment at the ``surface'' of the \hbindex{dynamo} region),
and~$B_{\mathrm{pole}}$ is the magnetic field strength at the surface
of the planet or cloud tops in the magnetic polar regions, which is
assumed to be the relevant region for the ECMI radiation.
Figure~\ref{fig:jupiter} provides an illustration of the exceptionally
sharp nature of the frequency truncation of the
\hbox{ECMI}.

\begin{figure}[htb]
\centering
\includegraphics[angle=-90,width=0.89\textwidth]{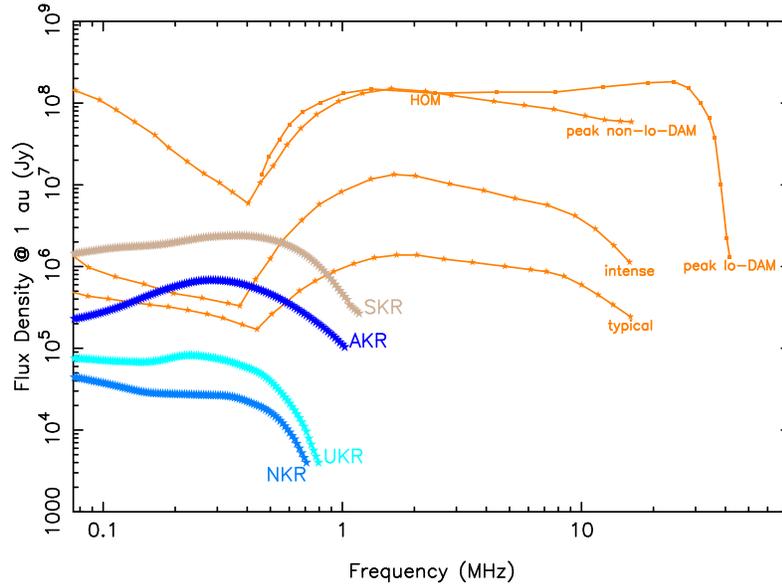}
\vspace*{-2ex}
\caption[Solar System Planetary Radio Spectra]%
{Radio spectra for the Solar System planets, scaled to a distance
  of~1~au.  For Jupiter, the hectometric emission (HOM), the
  decametric component linked to the moon Io (Io-DAM), and the
  decametric component not linked to Io (non-Io-DAM) are shown.  The
  intensities vary depending upon solar wind and internal
  magnetospheric conditions, and ``typical,'' ``intense,'' and
  ``peak'' values are shown, with ``peak'' occurring approximately 1\%
  of the time.  For the other planets, there is only a single
  contribution to the radio emission, which are the Saturnian
  kilometric radiation (SKR), the Earth's auroral kilometric radiation
  (AKR), the Uranian kilometric radiation (UKR), and the Neptunian
  kilometric radiation (NKR).  The Earth's AKR also is termed the
  terrestrial kilometric radiation (TKR).
  [Data courtesy of P.~Zarka.]}
\label{fig:radiosolarsystem}
\end{figure}

Figure~\ref{fig:radiosolarsystem} shows the (average) radio spectra for the
planets in the Solar System that sustain the \hbox{ECMI} \citep{z92}.
Jupiter, with a polar magnetic field strength at the cloud tops of
about~14~\hbox{G}, is clearly the most intense emitter, and the only
Solar System planet detectable from the ground \citep{bf55,fb56}.
With the other magnetic planets having much smaller magnetic
moments---the Earth's polar magnetic field strength is only
about~1~G---their maximum emission frequencies are below the
terrestrial ionospheric cutoff ($\sim 10$~MHz), which makes their
emissions unobservable from the ground.

The third conclusion is that, all other things being equal, more
intense emissions will be generated when more (supra-thermal)
energetic particles are available.  This conclusion follows from the
need to have an electron energy distribution that can feed energy into
the escaping radiation modes.  With more energetic particles, or a
more extreme supra-thermal distribution of energies, the ECMI can work
more effectively.
Extreme examples of solar wind control of the ECMI have been observed,
including a factor of~100 increase in the power of the Earth's ECMI
emissions from a factor of about~2 increase in solar wind velocity
\citep{gd81} and about a factor of~100 decrease in the power of
Saturn's ECMI emissions when Saturn moved into the trailing region of
Jupiter's \hbindex{magnetosphere} \citep{d83}.

A related aspect of Figure~\ref{fig:radiosolarsystem} concerns the
nature of Jupiter's radio emissions.  At higher frequencies ($\nu
\gtrsim 10$~MHz), Jupiter's decametric radio emissions (DAM) are a
combination of those related to the presence of its moon Io (so-called
Io-DAM) and those not related to Io
\citep[\hbox{non-Io-DAM},][]{gkh+02,2012P&SS...70..114H,2023arXiv230805541L}.  (There are
weaker contributions from Europa and Ganymede as well.)  A common
misconception is that Jupiter's radio emission results solely by the
presence of Io, and, by extension, a satellite is required for a
Jovian-mass planet to generate radio emission.

It is beyond the scope of this chapter, but the presence of a
planetary-scale magnetic field also may result in Ohmic dissipation
occurring in the planet's atmosphere as it moves through the stellar
wind of the host star.  The Ohmic dissipation would represent a heat
source, in addition to the stellar insolation, which could explain the
inflated radii of some ``hot Jupiters'' \cite[e.g.,][]{bs10,pmr10}.

\section{Extensions to and Predictions for Extrasolar Planets}\label{sec:radioexo}

Soon after the recognition that Solar System planets could be radio
emitters, \cite{fm74}, \cite{gg74}, \cite{yse77}, and \cite{wdb86} speculated about
and conducted searches for analogous emission from extrasolar planets.
If extrasolar planets host magnetic fields, it is reasonable to expect
them to generate radio emission via the ECMI as well, though the
challenges in detecting it are clear from simple considerations.  At
distances of at least $10^5$ times larger than for Solar System
planets (Figure~\ref{fig:radiosolarsystem}), the flux densities of extrasolar planets should be lower by
factors of at least $10^{10}$, though there also may be mechanisms
that would lead to (much) enhanced flux densities relative to what
such simple considerations might predict.

The flux density of an extrasolar planet depends upon the source
of available energy to the planetary \hbindex{magnetosphere}.
Five different input sources have been considered:
\begin{description}
\item[Stellar Wind Kinetic Energy]%
  The flux of protons of the host star's stellar wind incident on the
  planet's \hbindex{magnetosphere} provides a power input proportional
  to $\rho v^2$, for a stellar wind of density~$\rho$ and
  velocity~$v$.  This input energy source has been the most frequently
  considered one for extrasolar planets
  \citep{zqr+97,fdz99,ztrr01,lf+04,s05,gmmr05,g07,gpkmmr07,gzs07}

\item[Stellar Wind Magnetic Energy]%
  The flux of magnetic energy, or the electromagnetic Poynting flux,
  from the interplanetary magnetic field incident on the planet's
  \hbindex{magnetosphere} provides a power input proportional $B^2v$,
  for an interplanetary magnetic field strength~$B$ embedded in the
  stellar wind \citep{ztrr01,z07,gzs07,jcc08,sz11}.

\item[Stellar Coronal Mass Ejections]%
  The kinetic energy of a stellar coronal mass ejection (CME)
  impacting a planetary \hbindex{magnetosphere} provides power to the
  \hbindex{magnetosphere}, in a manner akin to the ``Stellar Wind
  Kinetic Energy'' above.  The distinction is that the kinetic energy
  of a CME is sufficiently large that the radio emission of the planet
  can be enhanced substantially relative to ``normal'' or quiet
  stellar conditions \citep{gd81,gmmr05,g07,gpkmmr07,gzs07}.

\item[Internal Magnetospheric Plasma Sources]%
  Because Jupiter's magnetic field rotates faster than Io orbits, a
  particle flux exists in the magnetic flux tube linking Io to
  Jupiter, and the ECMI is operative producing intense decametric
  radio emission (\hbox{Io-DAM}, Figure~\ref{fig:radiosolarsystem});
  as noted above, weaker contributions are produced by the flux tube
  footprints linking Europa and Ganymede as well.  This mechanism
  could result in strong radio emissions from a planet, even if it is
  sufficiently distant from its host star that the stellar wind
  pressure is relatively low.  Further, \cite{n11,n12} showed a planet
  with corotation-dominated \hbindex{magnetosphere}, such as that of
  Jupiter, orbiting a star with strong X-ray emission also could
  produce radio emission if there is enhanced coupling between
  planet's ionosphere, where the stellar X-ray photons deposit energy,
  and its \hbindex{magnetosphere}.

\item[Unipolar Interaction (Planet-Moon or Star-Planet)]%
  There could be equivalents of the Jupiter-Io system in which the
  interaction between a planet's \hbindex{magnetosphere} and its moon,
  a so-called exomoon, generate radio emission \citep{nsm14}.  
  Alternately, \emph{star-planet} systems with magnetic interactions
  could produce radio emission in an analogous manner.  In such a
  case, there would be a magnetic flux tube linking the planet and the
  star, and the ECMI would be operative in the \emph{stellar}
  \hbindex{magnetosphere} as the result of an orbiting planet,
  regardless of whether that planet has a magnetic field or not
  \citep{ztrr01,z06,z07,gpkmmr07,jcc08,sz11}.
  While this method could reveal the existence of a planet, it
  is not clear that this mechanism would provide much insight into the
  planet itself.
\end{description}
Generally, these models predict that close-in planets, especially
``hot Jupiters,'' should have more intense emissions, due to the
higher stellar wind loading of the \hbindex{magnetosphere}.
Amplification factors could be of order $10^3$, though, at very close
distances, within the closed \hbindex{magnetosphere} of the host star,
the ECMI mechanism may saturate rather than continue to increase
\citep{jcc08}.

A cautionary consideration regarding ``hot Jupiters,'' however, is
that they could be too close to their host stars for detectable radio
emission to be produced.  A star's stellar wind is a plasma, which has
its own plasma frequency.  If the stellar wind density near a ``hot
Jupiter'' is too high, its plasma frequency may be higher than the
cyclotron frequency of the planet's radio emission, and the planet's
radio emission would not be able to escape nor be detected \citep{gzs07}.

Further caution is required when applying simple scaling laws for
radio emissions produced by stellar wind interactions.
The solar wind-auroral radio emission connection may not be direct in
large corotating \hbindex{magnetospheres}.  Earth's convection-driven
\hbindex{magnetosphere} is especially sensitive to solar wind
pressure, and Earth's auroral kilometric radiation (AKR) can show a factor
of~100 increase in power for a factor of about~2 increase in solar
wind velocity \citep{gd81}.  However, Jovian aurora are driven by
currents that form in the co-rotating outer \hbindex{magnetosphere}
\citep{n11}, where the solar wind may impose only a secondary
controlling influence. There is a correlation of Jovian radio power
with the solar wind, but it is not as evident as the AKR case
\citep{gkh+02}.

As the stellar wind parameters strongly depend on the stellar age
\citep{2001ApJ...547L..49W,wmzl02,wmzlr05}, for those mechanisms that
depend upon energy input from the stellar wind, the age of the host
star also must be incorporated into predictions
\citep{s05,gmmr05,l+10}.  The radio flux of a planet around a young
star may be orders of magnitude higher than for a planet in an older
system.  Unfortunately, stellar ages are often poorly constrained, in
turn often leading to significant ranges in the predicted planetary
flux densities.

By the same token, if a planet is in an eccentric orbit, the effective
stellar wind density and velocity at the planet's
\hbindex{magnetosphere} will vary over the course of the planet's
orbit, in turn modulating the planet's emission \citep{gpkmmr07}.  In
the most dramatic cases, the modulation of the planet's radio emission
might approach a factor of~$10^3$ \citep{lsfb10}.  However, if the
planet's orbit is sufficiently eccentric, it may be carried into a
region where the \emph{stellar} wind plasma density is sufficiently
high that the considerations above concerning ``hot Jupiters'' become
relevant.

Regardless of the energy source powering the radio emission, the same
constraints of equation~(\ref{eqn:fmax}) apply for extrasolar planets
as for Solar System planets, namely, only those with sufficiently
strong magnetic fields will generate radio emission at a high enough
frequency to be detectable from the ground.  Estimating this frequency
for an extrasolar planet requires an estimate of the planetary
magnetic moment, which is often ill-constrained.  Two main approaches
have been adopted.  \cite{fdz99} and \cite{gzs07} assume the planetary
magnetic moment can be calculated by a force balance, and find a
planetary magnetic field that depends on the planetary rotation rate.
In contrast, \cite{rc10} assume the planetary magnetic moment to be
driven primarily by the energy flux from the planetary core. Thus,
they find no dependence on the planetary rotation rate; however, they
obtain stronger magnetic fields and more favorable observing
conditions for young planets.  \cite{do11} considered the specific
case of terrestrial planets.  They found that anomalously strong
fields ($3\times$ larger than the most optimistic prediction) are
required for emission at frequencies above the Earth's ionospheric
cutoff; furthermore, the expected flux levels are very low.

Finally, in the interest of completeness, planets in more ``exotic''
environments have been considered as possible radio emitters.  These
environments include planets around pulsars \citep{mcvf23},
terrestrial planets around white dwarfs \citep{ww05}, planets around
evolved cool stars \citep{igl10,fsm+15}, planets around T~Tauri stars
\citep{voj-pg10}, and even interstellar ``rogue planets,'' i.e.,
planets not bound to a star \citep{v11}.  While there nature is not
yet clear, Jupiter-Mass Binary Objects (JuMBOs) may be planets
\citep{2023arXiv231001231P}, and at least one such JuMBO has been
detected at radio wavelengths \citep{rlz24}.

\section{Planetary Magnetic Fields and Interiors}\label{sec:interior}

The detection and measurement of extrasolar planetary magnetic fields,
whether in the context of discovering new extrasolar planets or
observing known extrasolar planets, could provide constraints on the
thermal states, compositions, and dynamics of extrasolar planetary
interiors.
The mass-radius diagram can provide some constraints for models of
planetary interiors, but, because the same bulk density can be
obtained by different admixtures of constituents (iron vs.\ silicates
vs.\ volatiles), there are considerable degeneracies
\citep[e.g.,][]{Rogers&Seager2010ApJ,2014PNAS..11112622S,2014ApJ...792....1L,sjr+17}.
Further, even for a planet with a fixed bulk composition, its location
on the mass-radius relation may change over time as the planet's
thermal state evolves \citep{2020A&A...638A.129N}.  It may even be the
case that the presence of a magnetic field may help determine a
planet's location in the mass-radius relation by shielding its
atmosphere from erosion \citep{2019MNRAS.490...15O}.  (See also the
section below, ``Planetary Magnetic Fields and Habitability.'')

The most simple approach to imposing constraints on a planet's
interior is that a planetary-scale magnetic field implies that there
must be an electrically-conducting region within a planet.  More
sophisticated approaches include using the period of rotation
determined from the magnetic field, or from radio emission linked to
the magnetic field, as an input to interior structure models, as
\cite{1991Sci...253..648H} did for Neptune.  There likely are rich
opportunities to explore for using detections of extrasolar planetary magnetic
fields to constrain the structure of planetary interiors
\citep[e.g.,][]{2013ApJ...768..156Y}.

Emerging approaches to provide additional constraints on a planet's
interior structure are to assume that the composition of the planet is
similar to that of its host star, at least for the refractory elements
\citep{2015A&A...577A..83D,2017A&A...597A..37D,2017ApJ...850...93B,2021PSJ.....2..113S,2023ApJ...944...42U},
or can be inferred from atmospheric abundances
\citep[e.g.,][]{2023MNRAS.523.6282B,gsjr23}.
The technique of linking the host star-planet compositions likely
requires high-precision spectroscopic measurements of the host star in
order to obtain sufficient constraints on a planet's composition
\citep{2019MNRAS.482.2222W}.  Moreover, there also are recent examples
of planets that appear to have compositions discrepant, sometimes
substantially so, from their host stars
\citep[e.g.,][]{bxa+23,swz+23}.  There is even the possibility
that a planet's composition may reflect not only how far from its host
star that it formed (radial location within the protoplanetary disk),
but also azimuthal variations or other variations (e.g., temperature)
within the protoplanetary disk \citep{kkb+23,2023arXiv230711817V}.
Nonetheless, there also may be opportunities to model the likelihood
of a planet generating a magnetic field using its host star
metallicity as an additional constraint.

The technique of inferring the interior composition from the
atmospheric composition typically uses spectroscopic observations of a
planet's atmosphere to infer the atmospheric (or envelope)
composition.  There is considerable interest in this technique with
the advent of the \textit{JWST}.  This technique depends crucially on
the extent to which the atmosphere or envelope is able to mix
homogeneously with lower layers (such as a mantle in the case of a
rocky planet).  As discussed by \cite[and references
within]{2023MNRAS.523.6282B}, even within the Solar System, the
results of the Juno mission at Jupiter and the \textit{Cassini}
mission at Saturn provide cautionary notes about this approach.

Measurements of, or constraints on, the strengths of extrasolar
planetary magnetic fields, even for a small number of extrasolar
planets, would be valuable from two, complementary perspectives.
First, the presence of a magnetic field requires the planet to support
an internal dynamo, which in turn requires some internal region of the
planet to support convection.  Considering a planet with a given mass
and radius, only a limited set of bulk compositions, thermal states,
and internal pressure profiles may enable a dynamo region to persist \citep[e.g.,][]{2013ApJ...768..156Y}.

Second, detecting planetary magnetic fields may enable new insights
about the planetary \hbindex{dynamo} process itself.  The wide variety
of magnetic fields observed for Solar System planets results in data
starvation for models and has confounded efforts to develop a
comprehensive description of planetary \hbindex{dynamos}.  Having more
examples of planetary magnetic fields on which to test models may
provide better information on the planets in the Solar System, much
like how the diverse nature of planetary systems have provided
insights into the mechanisms by which planets form and evolve.

The constraints that a magnetic field measurement would provide depend
upon the class of the planet.

\subsection{Gas-Giant Planets}\label{sec:interior.gas}

For gas-giant planets (i.e., Jupiter mass), the equation of state of
hydrogen is known sufficiently well to be confident that it undergoes
a transition to a metallic state in a gas-giant planet's interior
\citep{2002ARA&A..40..103H,hmr20}.  Further,
\citet{2018AJ....155..178B} argues that rotation rates of gas giants
planets being well below their breakup velocities is consistent with
magnetic braking of the planet due to its magnetic field being coupled
to the (ionized) circumplanetary disk.
Consequently, all gas-giant
planets are expected to sustain planetary-scale magnetic fields, but
there might be a wide range of magnetic field strengths.
For instance, estimates for the polar magnetic field strength of
HD~209458~b, achieved with a variety of different (model-dependent)
methods, have ranged over as much as two orders of magnitude
\citep{s-l04,bs10,fsy+10,2010ApJ...709..670E,kslmrbf21}, from much smaller than that of Jupiter to much
larger.

The \emph{absence} of a magnetic field in a gas-giant planet would be
the more consequential result.

\subsection{Ice-Giant Planets}\label{sec:interior.ice}

Based on the Voyager~2 measurements of the magnetic fields of Uranus
and Neptune, the standard explanation is that the dynamo regions of
ice-giant planets are in ionic layers, located at roughly 70\% of the
planetary radii \citep{2004Natur.428..151S,2011ApJ...726...15H}.
These ionic layers would contain the dominant volatiles of ice
giants---water, ammonia, methane, or some combination of all.

A wide variety of different internal compositions and
structures have been explored.
\cite{2013ApJ...768..156Y} modeled the dynamo regions of planets with
masses ranging from terrestrial mass to ice giants, with a focus on
the geometry of the dynamo region, which could be characterized by the
thickness of the dynamo region.  They found that the thickness of the
dynamo region varied, depending upon the mass, temperature, and
composition.  In turn, the thickness of the dynamo region produced
substantially different magnetic field morphologies, with largely
axial and dipolar magnetic fields resulting from thick dynamo regions
and non-axisymmetric and multipolar magnetic fields resulting from
thin dynamo regions.

The composition of the \hbindex{dynamo} region of an ice giant may be
more complex than can be inferred from a single spacecraft flyby
(i.e., from the Voyager~2 results).
\cite{rbh+21} demonstrated, using a sample at pressures and
temperatures likely achieved in the interior of ice giants, that
ammonia may undergo a phase transition to a plasma or metallic state.
In this state, they found that ammonia could have an electrical
conductivity up to an order of magnitude larger than that of water,
and they concluded ice giant magnetic fields might be produced in the
regions that are the most ammonia-rich.
Similarly, \citep{ooh24} find that the electrical conductivity of
water at the temperatures and pressures in the interior of an ice
giant does not appear sufficient to sustain a \hbindex{dynamo}.
Finally, \cite{n16,n17} has argued that metallic hydrogen in the
interfaces between the envelopes of Uranus and Neptune and their lower
layers is responsible for generating their \hbindex{dynamos}.

Observationally, there are two clear directions to pursue.  Within the
Solar System, a future Uranus Orbiter, such as that recommended by the
\cite{owl} report from the Planetary Science \& Astrobiology Decadal
Survey in the United States, could obtain a much
higher fidelity characterization of the Uranian magnetic field.
However, the Uranus Orbiter would provide measurements for only a
single ice giant, and a potentially unrepresentative one.  With both
its high obliquity and low thermal flux, the extent to which Uranus'
interior structure can be taken to be representative of all ice giants
is yet to be determined.

A complementary approach is to obtain magnetic field measurements for
even a modest sample of extrasolar ice giants.  A testable hypothesis
motivating such measurements could be that ice giants produce magnetic
fields with a broad range of strengths, with Uranus and Neptune at the
low end of this range.  This broad range of magnetic field strengths
could result from various origins or evolutions.  For instance, ice
giants forming at different distances from their host stars might have
different bulk compositions, which could affect whether a water-rich
or ammonia-rich internal layer generates the dynamo.  Similarly, ice
giants might produce a range of magnetic field topologies, ranging
from the complex multi-polar topologies of Uranus and Neptune to more
dipolar topologies characteristic of Jupiter and Earth.  At a
distance, dipolar topologies would appear to have larger magnetic field
strengths than multi-polar topologies.

\subsection{Super Earths}\label{sec:interior.super}

Because the Solar System contains no sub-Neptunes nor
super-Earths, there is an opportunity for discovery should magnetic
fields be able to be measured for these classes of planets.  Indeed,
\cite{csw+21} have measured various properties of samples under the
pressures and temperatures expected in the interiors of super-Earths.
They find that the higher temperatures and pressures in super-Earths,
relative to terrestrial-mass planets, could result in different
mineralogical properties, including lower viscosities.  A potential
consequence, they speculate, could be that super-Earths would have an
increased ``electromagnetic coupling between the core and the mantle,
enhancing convection and heat transport out of the core and affecting
the strength and expression of the magnetic field.''

\subsection{Terrestrial-Mass Planets}\label{sec:interior.earth}

The dynamos, if not interior structures, of terrestrial-mass planets
can be expected to exhibit a large range.  In the Solar System, Earth
exhibits a strong, dipolar magnetic field, while Venus produces no
(current) planetary magnetic field, even though the two planets are
nearly ``twins.''

\begin{figure}
  \centering
  \includegraphics[angle=-90,width=0.97\textwidth]{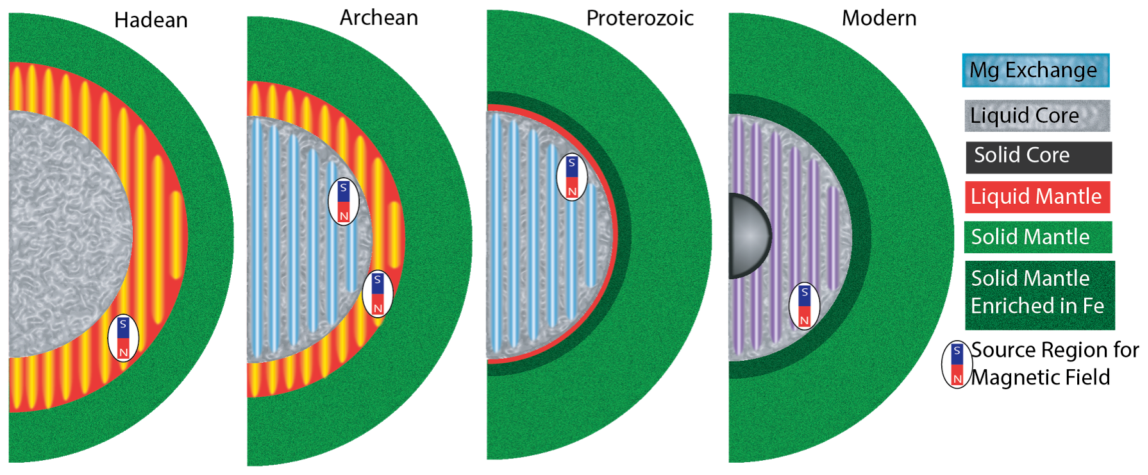}
  \vspace*{-5ex}
  \caption[Planetary Dynamo Evolution]%
    {Illustration of how the source region of the Earth's
    magnetic field may have changed as the Earth has cooled over
    geological time scales.  The different source regions potentially
    would have generated magnetic fields of different strengths, with
    concomitant implications for the Earth's radio emissions.  Similar
    processes may occur for other terrestrial-mass planets. (Credit: D.~Stegman)}
  \label{fig:terrestrialsource}  
\end{figure}

\cite{2020A&A...638A.129N} have shown that variations in the relative
fractions of silicates and iron within terrestrial-mass planets could
produce substantially different temperature-pressure profiles.  While
they do not extend their analysis to modeling the generation of a
dynamo, differences in the interior state reasonably can be expected
to result in different magnetic field strengths, topologies, or both.  
Indeed, estimates of the Earth's paleomagnetic field show variations
(in the median) of approximately a factor of a few over the past
3500~Myr \citep[e.g.,][]{2022GeoRL..4900898B}, suggesting that the
dynamo evolved (substantially) over geologic time as its interior
properties evolved.  Further, \cite{zs13} and \cite{bsz20} illustrate how the Earth's
dynamo may have shifted from its mantle to its core as its interior
cooled and the inner core began to solidify
(Figure~\ref{fig:terrestrialsource}).  If the dynamo has evolved over
geological time scales, the Earth's radio emissions may have had
considerably different intensities as well.

Observationally, there is  considerable information on the magnetic
fields of terrestrial-mass planets within the Solar System.  Finding
and characterizing terrestrial-mass planets will be a significant
focus for the coming decades, as discussed in the \cite{astro2020}
report from the recent Decadal Survey in Astronomy \& Astrophysics
conducted in the United States.
Determining magnetic field strengths of terrestrial-mass planets not
only may yield information about their interior structures, in a
manner analogous to ice giants, but also be relevant for their
habitability, as the next section discusses.

\section{Planetary Magnetic Fields and Habitability}\label{sec:habitable}

A challenging, yet intriguing, possibility is that the detection of
extrasolar planetary magnetic fields may provide information about the
potential \hbindex{habitability} of terrestrial planets, or help
explain why some terrestrial planets are not inhabited.  As summarized
in the \cite{exostrategy}, a planetary magnetic field could protect
the (secondary) atmosphere (and potentially surface) of a terrestrial
planet from cosmic rays and the effects of intense stellar flares and
eruptions (coronal mass ejections, CMEs).  \citep[See also][]{2019Sci...364..434S}.

There is a rich literature regarding how a planet's magnetic field
might protect its (secondary) atmosphere from erosion by its host
star's stellar wind or its surface from the effects of high-energy
particles or both
\citep{griessmeier2004,2005AsBio...5..587G,2007SSRv..129..279D,2008SSRv..139..399L,2009Icar..199..526G,griessmeier2010,2013Icar..226.1447D,2015A&A...581A..44G,2016GGG....17.1885F,2017ApJ...844L..13G,2019MNRAS.490...15O}.
The essential concept is that, if a planet's magnetic field is
sufficiently strong, and the kinetic and magnetic energies of the host
star's stellar wind are not too strong, the energetic particles in the
host star's stellar wind are deflected (by $\mathbf{v} \times
\mathbf{B}$ forces) before they reach the planet's atmosphere.
This concept has been extended to consider the possibility that, even
if a planet itself is not habitable, its \hbindex{magnetosphere} could
contribute to protecting a moon, thereby contributing to the moon's
habitability \citep{gbd21}.

This potential importance was identified in \cite{astro2020} by the
Science Priority Question~2, ``What are the Properties of Individual
Planets and Which Processes Lead to Planetary Diversity?''  with the
secondary questions, Question~2b, ``How Does a Planet’s Interior
Structure and Composition Connect to Its Surface and Atmosphere?'' and
Question~2d, ``How Does a Planet’s Interaction with Its Host Star and
Planetary System Influence Its Atmospheric Properties over All Time
Scales?''

There have been efforts to assess the role of the Earth's
magnetosphere in determining its habitability \citep{vbsrzp23} and
even speculation that changes in the Earth's \hbindex{paleomagnetosphere}
may have contributed to substantial changes in the evolution of life
\citep[e.g.,][]{mlbl16}.
This expectation of atmospheric protection is has been extended to be
included in models that attempt to assess or predict the likely
habitability of extrasolar planets
\citep{2017MNRAS.471.4628R,2019A&A...630A..52R,2019MNRAS.485.3999M}.

The possibility of a stellar wind eroding completely, or nearly so, a
planet's atmosphere is particularly acute in the case of a planet
orbiting an M~dwarf, for two reasons.  First, the lower stellar
insolation means that the planet must orbit (much) closer to the star
in order to be in the traditional habitable zone where the temperature
is high enough that liquid water could exist on the planet's surface.
Second, M~dwarfs can have intense stellar wind activity.  Notably, the
closest potentially habitable planet may orbit Proxima Centauri~b,
emphasizing these concerns \citep{2016A&A...596A.111R,2017ApJ...844L..13G}.
Additionally, the extent to which a planet's atmosphere is exposed to
high-energy particles, either from its host star or Galactic cosmic
rays, may affect its chemistry \citep{2016A&A...587A.159G}.

Dramatic evidence of the effects on a planetary atmosphere without
magnetic shielding, and exposed to the effects of a stellar wind, were
provided by Mars Atmosphere and Volatile Evolution (MAVEN)
observations showing erosion of the Martian atmosphere when it was
struck by a solar CME \citep{jgl+15}.
However, a more nuanced view notes that the Earth's
\hbindex{magnetosphere} presents a larger obstacle in the solar wind
as compared to the ionospheres of Venus or Mars, allowing for more
interactions and greater rate of atmospheric loss
\cite{2018MNRAS.481.5146B}.  Recent reviews, motivated by space
physics observations within the Solar System, have emphasized that
Earth has both a stronger magnetic field than either Venus or Mars and
a \emph{larger} atmospheric loss rate
\citep{2020JGRA..12527639G,2021SSRv..217...36R}; of particular note is
Figure~5 of \cite{2021SSRv..217...36R}.
Even for a larger planet, interactions between its
\hbindex{magnetosphere} and the host's star stellar wind can inject
energy into its atmosphere, at least inflating if, and potentially
increasing its mass loss \citep{l13}.

A complication regarding the atmospheric shielding effects of a
planetary magnetic field is that a magnetic field is only effective in
shielding energetic charged particles.  Other processes, such as
photoevaporation from the host star's soft X-ray and UV emission or
core-powered mass loss, also may contribute to atmospheric loss.

\section{Future Steps}\label{sec:future}

Since the first edition of the \textit{Handbook of Exoplanets}, there
have been a number of notable improvements in capabilities and new
searches, albeit there remains no current, unambiguous detection of
radio emission from an extrasolar planet
(Figure~\ref{fig:radio_limits}).
There are two, complementary avenues that hold promise for the future.
New (ground-based) capability is emerging below~50~MHz, as exemplified
by the observations of HD~80606~b \citep{2020A&A...644A.157D}, and the
potential for future space-based missions is increasing.

\begin{figure}[tbh]
\centering
\includegraphics[width=0.47\textwidth]{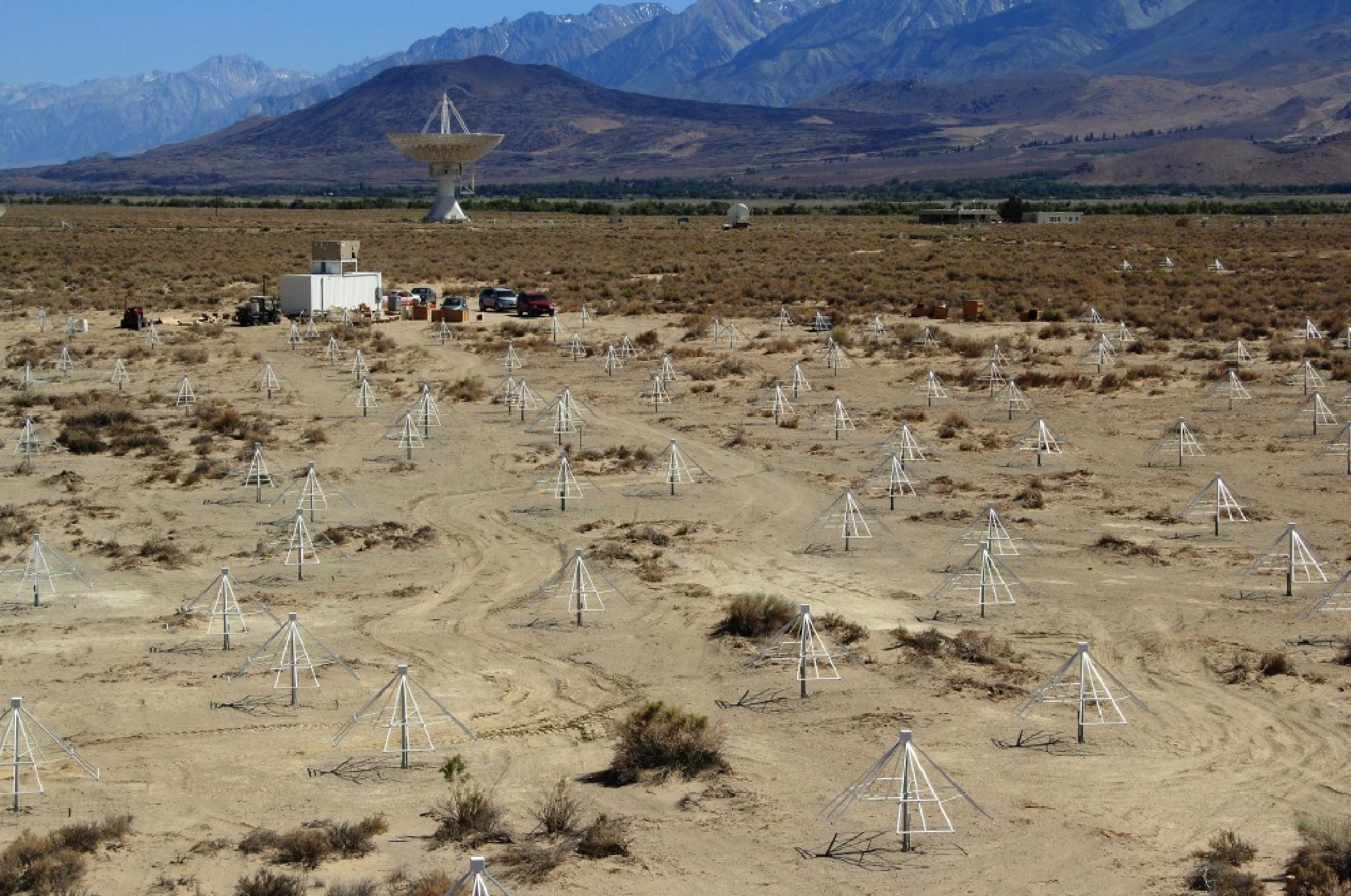}\hfil%
\includegraphics[width=0.47\textwidth]{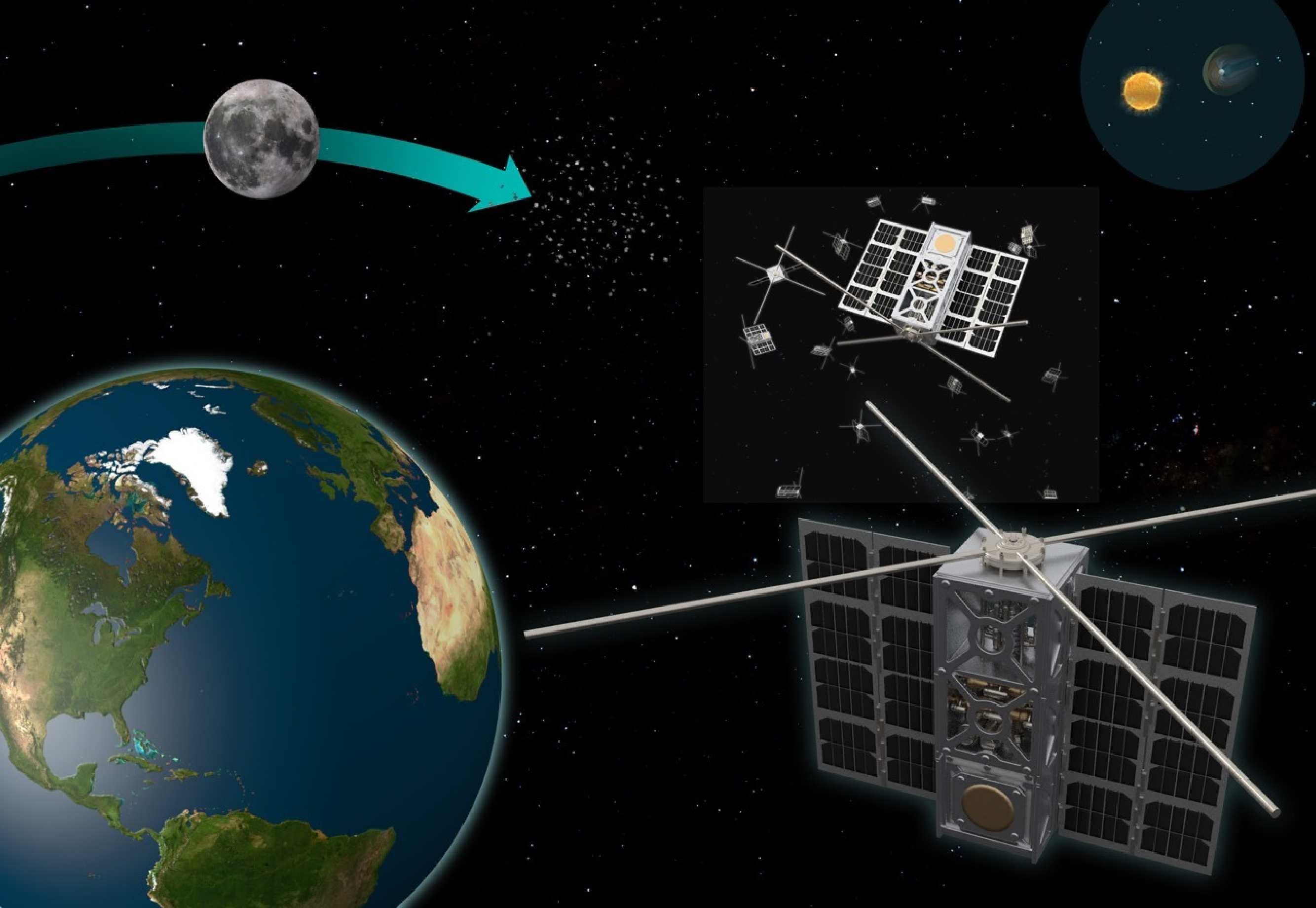}
\vspace*{-1ex}
\caption[Exoplanet Radio Telescopes]%
{(\textit{Left}) Long Wavelength Array at Owens Valley Radio
Observatory (\hbindex{LWA-OVRO}), one of the new capabilities for
searching for and observing the extrasolar planetary radio emission and a possible model for future space-based observations.
The signals from all of the individual dipole antennas are transmitted
to a central location for processing, producing an effective aperture
equivalent to the approximate maximum separation between the antennas
($\approx 300$~m).  (In the background are other antennas at the Owens
Valley Radio Observatory.)  The low-band antennas of the Low Frequency
Array (\hbindex{LOFAR}/LBA) are more closely packed in individual ``stations''
or groups, but the stations are widely distributed across the
Netherlands and Europe.
(\textit{Right}) Artist's impression of a possible future space-based
radio telescope for observing radio emission from extrasolar planets.
Space-based observations are likely to be required for detecting
planets with magnetic fields weaker than about~10~\hbox{G}.  This
architecture is analogous to that of the \hbindex{LWA-OVRO} and
\hbox{\hbindex{LOFAR}/LBA}, with each small spacecraft carrying a single dipole
antenna.  Signals from the individual spacecraft would be combined to
form the synthetic aperture.}
\label{fig:radiotelescopes}
\end{figure}

There are three primary (ground-based) telescopes enabling future
searches and observations below~50~MHz, \hbindex{LOFAR}, the Long
Wavelength Array at the Owens Valley Radio Observatory
(\hbindex{LWA-OVRO}), and \hbindex{NenuFAR}.
\begin{description}
\item[The Low Frequency Array]%
  \hbindex{LOFAR} can observe from~10~MHz to~90~MHz \citep[and at
    higher frequencies, ][]{lofar}.  There is a plan to upgrade its
  low-band antennas (LBAs) in order to provide increased sensitivity
  \citep{lofar2.0}, and there is an increasing emphasis on improving
  the calibration of the telescope at these low frequencies, which
  must contend with the Earth's ionosphere.

  \item[\hbindex{LWA-OVRO}] The Long Wavelength Array at the Owens
    Valley Radio Observatory (\hbindex{LWA-OVRO}) can observe between
    about~20~MHz and~80~MHz (Figure~\ref{fig:radiotelescopes}).
    Building on the lessons from the first station of the LWA
    \citep[LWA1,][]{tek+12}, the \hbindex{LWA-OVRO} has enhanced
    imaging capabilities, which provide a higher sensitivity.  The
    \hbindex{LWA-OVRO} has been undergoing an expansion-upgrade
    project and, while still in its commissioning phase, initial
    observations are promising.
  
  \item[\hbindex{NenuFAR}] is a substantially enlarged station (``super
    station'') of \hbindex{LOFAR}-like antennas that can be used as part
    of \hbindex{LOFAR} or as a stand-alone telescope \citep{nenufar}.
\end{description}

A crucial aspect for searching for radio emission from extrasolar
planets is that both \hbindex{LOFAR} and \hbindex{LWA-OVRO} provide
wide-field capabilities, enabling surveys of the sky to be conducted
\citep[e.g.,][]{lotss}. 
Such surveys have resulted in the discovery of radio emission from
low-mass stars, possibly indicative of star-planet interactions
\citep{vcs+20}.
Further, because planetary radio emission is expected to be weak, the
most likely first detections will be of planets in the solar
neighborhood.  Their host stars are distributed widely on the sky, so
a wide-field capability enables a radio telescope to monitor multiple
stars simultaneously.  Such long-term monitoring capability is
particularly important given that there may be (likely are) geometric
effects that result in planets illuminating the Earth only during
specific phases of their orbits.

In this context, motivated in part by the Habitable Worlds Observatory
(HWO) concept, over the next few years, a much more complete census of
extrasolar planets in the solar neighborhood is expected
\citep[e.g.,]{ms23}, from a variety of both ground- and space-based
efforts.  A possible consequence is that, within the volume for which
it is likely that radio emissions could be detected, there will be few
extrasolar planets discovered initially by their radio emissions
because other techniques will discover them first.  Conversely, this
census of nearby extrasolar planets will provide specific targets for
radio telescopes.

Finally, both \hbox{LOFAR} and \hbindex{LWA-OVRO} are making
considerable efforts to calibrate their polarization responses.  As
the {\hbindex{electron cyclotron maser instability}} (ECMI) naturally
produces circularly polarized radiation, and few other sources produce
high levels of circular polarization, searches for
circularly-polarized sources could be a particularly fruitful approach
to discovering the radio emissions of extrasolar planets.
Indeed, at least one such survey for circularly polarized sources is
underway using \hbindex{LOFAR}, the V-LoTSS \citep{v-lotss}.

While the focus of this chapter has been on the
detection and study of extrasolar planets at radio wavelengths, a
related aspect is any potential connection between gas-giant planets
and brown dwarfs.  Both are (degenerate) sub-stellar objects capable
of generating large-scale magnetic fields, and scaling laws suggest
that there should be a continuum of magnetic field strengths between
planets and brown dwarfs
\citep{Christensen10,Stevenson2010SSR,Schubert&Soderlund2011PEPI,chr09}.
Brown dwarfs have been detected at radio wavelengths
\citep[e.g.,][]{bbb+01,rw16,lmrhlw16,pv23,rpm+23}, and, at least for
some objects, their radio emissions have been confirmed to be due to
an ECMI in their auroral regions \citep{hadblg08,khpebbs16,khpsb18}.
Further, the ultracool dwarf LSR~J1835+3259, which may be either a
brown dwarf or a low-mass star, supports radiation belts akin to those
of Jupiter \citep{kmvs23,2023arXiv230306453C}, providing additional
support for the hypothesis that the generation of magnetic fields
within brown dwarfs and giant planets likely share similarities.  To
date, though, brown dwarfs have been detected mostly at frequencies
$\nu \gtrsim 5$~GHz, implying magnetic field strengths $B \gtrsim
1\,\mathrm{kG}$; efforts to detect them at lower frequencies largely have been
 unsuccessful \citep{jolkm11,bhn+16}.  Further study of brown
dwarfs may provide clues to guide the discovery of radio emission from
extrasolar planets.

This chapter also touches only lightly upon the larger topic of
\textit{extrasolar space weather} \citep{och18,lorentz}, though it clearly is
linked to the radio emission of extrasolar planets.
  
It has been beyond the scope of this chapter, but the on-going Juno
mission continues to provide insights about Jupiter's interior
structure and its radio emissions
\citep[e.g.,][]{ljh+22,2023arXiv230805541L}, and will continue to do
so until at least 2025 \citep{juno}.
Complementing the Juno science investigation, the Jupiter ICy moons
Explorer (JUICE) is scheduled to arrive into the Jupiter system
in~2031, where it also will study the Jovian interior and
\hbindex{magnetosphere} \citep{2013P&SS...78....1G}.
Ice giants may be among the most numerous of extrasolar planets, but
there has never been a dedicated mission to either Uranus or Neptune,
only a single fly-by of both planets by the Voyager~2 spacecraft.
The recent \cite{owl} Decadal Survey report in the United States
recommends that a Uranus Orbiter \& Probe mission be a future NASA
Flagship mission, with the science motivation being in part to
understand Uranus' interior and \hbindex{magnetosphere}.

If the Solar System planets are any guide,
Figure~\ref{fig:radiosolarsystem} emphasizes that observations must be
able to be conducted below~10~MHz, i.e., the approximate plasma
frequency of the Earth's ionosphere.  The concept of a space-based
radio telescope is not new \citep[e.g.,][]{fhr67}, and there have been
the initial demonstrations of the capability to conduct radio
astronomical observations from space.  The first Radio Astronomy
Explorer (RAE-1) was in an Earth orbit and made the first measurements
of the Galaxy's spectrum between~0.4 and~6.5~MHz \citep{abcsw69} while
the second Radio Astronomy Explorer (RAE-2) was in a lunar orbit and
observed between~25~kHz and 13~MHz \citep{akngw75}.  Jupiter's radio
emission was detected with both spacecraft \citep{1974ApJ...194L..57D,1977JGR....82.1256K}.
The Earth's AKR has been studied by simple space-based telescopes
including a single-element interferometer consisting of the ISEE-1 and
ISEE-2 spacecraft \citep{bgcs86} and a time-difference-of-arrival
(TDOA) analysis with the Cluster spacecraft \citep{mgc04}.

There have been initial descriptions and proposals of concepts
for radio astronomy arrays of small spacecraft, notably including the
Astronomical Low Frequency Array (ALFA) mission concept \citep{alfa},
and ``CubeSat''-based arrays \citep{bljsfa13}, for which the detection
and study of extrasolar planets was either a part of the science
mission or the prime science mission.  Most of the attention for
future space-based radio telescopes has been focussed on
constellations of small spacecraft (Figure~\ref{fig:radiotelescopes}),
which are straightforward analogies to how current radio telescopes in
this frequency range are realized.

An initial realization of such a space-based radio telescope will be
the Sun Radio Interferometer Space Experiment
\citep[\hbox{SunRISE},[]{klr-wln22}, an array of six CubeSats.  The
primary science focus for SunRISE is determining the locations of
solar radio bursts, and, with only six antennas, it will not have the
sensitivity to detect the radio emissions from extrasolar planets.
However, as a pathfinder to future, larger constellations, it likely
will provide valuable lessons.

Moreover, in a microgravity environment, it may be possible to
contemplate much larger single apertures than is possible on the
Earth.  The two approaches have different strengths and weaknesses
\citep{kiss}---a constellation of small spacecraft may generate
infeasible data rates while there is little experience in constructing
extremely large structures ($> 100\,\mathrm{m}$) in space.

Combined with new
capabilities at low radio frequencies, the study of extrasolar planets
at radio wavelengths remains a promising field.

\begin{acknowledgement}
It is a pleasure to thank many colleagues for their discussions,
speculations, and multiple explanations on the topic of
(exo-)planetary radio emission, most notably B.~Farrell, P.~Zarka,
J.~Callingham, 
J.~M.~Griessmeier, G.~Hallinan, E.~Shkolnik, D.~Stevenson, J.~Turner, K.~Weiler,
and L.~Henning and the students at Thomas Jefferson High School for
Science \& Technology.  The author acknowledges the ideas and advice from the
participants in the ``Planetary Magnetic Fields: Planetary Interiors
and Habitability'' workshop organized by the W.~M.~Keck Institute for
Space Studies.
This research has made use of NASA's Astrophysics Data System.
Part of this research was carried out at the Jet Propulsion
Laboratory, California Institute of Technology, under a contract with
the National Aeronautics and Space Administration.
Some of this material is pre-decisional information and for planning
and discussion only.
\end{acknowledgement}

\section*{Cross-References}

\begin{itemize}
\item 
  Planetary Interiors, Magnetic Fields, and Habitability
  
\item
  Future Exoplanet Research: Radio Detection and Characterization

\item
  Star-Planet Interactions in the Radio Domain: Prospect for Their
  Detection

\item 
  Magnetic Environment of the Planets

\item 
  Radio Emission from Ultracool Dwarfs

\item 
  Pulsar Timing as an Exoplanet Discovery Method

\item
  The Solar System as a Benchmark for Exoplanet Systems Interpretation

\item 
  Factors Affecting Exoplanet Habitability

\item 
  Interiors and Surfaces of Terrestrial Planets and Major Satellites

\item 
  Internal Structure of Giant and Icy Planets: Importance of Heavy Elements and Mixing
\end{itemize}

\end{document}